\documentclass[conference]{IEEEtran}
\IEEEoverridecommandlockouts

\usepackage{amsmath,amssymb,amsfonts}
\usepackage{lscape}
\usepackage{epsfig}
\usepackage{graphicx}
\usepackage{tabularx}
\usepackage[table]{xcolor}
\usepackage{textcomp}
\usepackage{booktabs}
\usepackage{textcomp}
\def\BibTeX{{\rm B\kern-.05em{\sc i\kern-.025em b}\kern-.08em
    T\kern-.1667em\lower.7ex\hbox{E}\kern-.125emX}}

\begin{document}

\title{A Novel Design Method for Digital FIR/IIR Filters Based on the Shuffle Frog-Leaping Algorithm \\
\thanks{This is the final manuscript (post-print) accepted and published in the Proceedings of the 27th. European Signal Processing Conference (EUSIPCO 2019), Coru$\tilde{n}$a, Spain, pp. 1-5, 2-6 Sept. 2019. ISBN: 978-9-0827-9703-9, ISSN: 2076-1465. DOI: https://doi.org/10.23919/EUSIPCO.2019.8903129. (c) 2019 IEEE.}
}

\author{
\IEEEauthorblockN{Daniel Jim\'enez-Galindo}
\IEEEauthorblockA{Universidad de Valladolid\\
Valladolid, Spain \\
danieljiga@gmail.com}
\and
\IEEEauthorblockN{Pablo Casaseca-de-la-Higuera}
\IEEEauthorblockA{Dept. Signal Theory and Communications \\
Universidad de Valladolid\\
Valladolid, Spain \\
casaseca@lpi.tel.uva.es}
\and
\IEEEauthorblockN{Luis M. San-Jos\'e-Revuelta}
\IEEEauthorblockA{Dept. Signal Theory and Communications \\
Universidad de Valladolid\\
Valladolid, Spain \\
lsanjose@tel.uva.es}
}

\maketitle

\begin{abstract} The design of both FIR and IIR digital filters is a multi-variable optimization problem, where traditional algorithms fail to obtain optimal solutions. A modified Shuffled Frog Leaping Algorithm (SFLA) is here proposed for the design of FIR and IIR discrete-time filters as close as possible to the desired filter frequency response. This algorithm can be considered a type of memetic algorithm. In this paper, simulations prove the obtained filters outperform those designed using the traditional bilinear Z transform (BZT) method with elliptic approximation. Besides, results are close to, and even slightly better, than those reported in recent bio-inspired approaches using algorithms such as particle swarm optimization (PSO), differential evolution (DE) and regularized global optimization (RGA).
\end{abstract}


\section{Introduction}

Digital filters are essential components in every signal processing system as they enhance valuable information, either by separating it from other unwanted signals or by attenuating the noisy components found in the  raw signal.

Two types of digital filters are studied in this paper: Finite Impulse Response (FIR) and Infinite Impulse Response (IIR) digital filters. FIR filters produce an output that is only based on current and past inputs, ensuring stability. On the other hand, IIR  filters produce an output based not only on its input, but also on the previous values of the output. As a consequence, stability is not guaranteed and is a desirable characteristic when designing them.

IIR digital filters imply about twice as much coefficients as the FIR filters, resulting in a better frequency response for the same filter order, nevertheless increasing the computational load of the proposed memetic algorithm. In addition, the search space must be limited to ensure the stability of the filter, introducing a number of checks that slow down convergence speed.

Many conventional methods allow the design of digital filters, but they only offer a limited range of parameters to be tuned-up in order to control the impulse response of the filter \cite{Wai09}. In addition, several optimization algorithms have already been used for digital filter design, but they suffer from a great limitation as they frequently get stuck in local minima, failing to achieve an optimal solution \cite{Krasnogor06}.

One of the first bio-inspired algorithms proposed for filter design used Genetic Algorithms (GAs) \cite{Back97}. These are based on the theory of evolution formulated by Charles Darwin and are quite simple to implement. Genetic algorithms established the basis for more recent memetic algorithms (MAs), which introduce the concept of {\em meme} \cite{Krasnogor06}, alluding the capability of each individual to communicate with the others.

The here proposed memetic algorithm is based on the traditional {\em Shuffled Frog-Leap Algorithm} (SFLA) initially described in \cite{Aggarwal04}. The algorithm relies on the analogy of a pond populated by frogs leaping to different positions that represent possible solutions to the studied problem. Various modifications have been applied so that the algorithm tackles the filter design challenge in a more efficient way.

Different techniques have been introduced to improve the basic algorithm such as the one mentioned in \cite{Aggarwal04}, which aims at ensuring a good population diversity, avoiding premature convergence and taking advantage of diversified search. These desired characteristics imply a more complex algorithm (compared to the standard one) that focuses on the quality of the final solution while sacrificing speed and computational resources.

The rest of the paper is organized as follows:  Section \ref{litrev} contains the literature review focused on bio-inspired algorithms. Section \ref{sec_theoretical} describes the mathematical framework for IIR and FIR filter design. Section \ref{method} outlines the main characteristics of the proposed method. Section \ref{sec_results} presents the results obtained by the studied SFLA together with a comparison with those obtained by both classical and bio-inspired strategies. Finally, conclusions are gathered in Section \ref{conclusion}.

\section{Literature review}
\label{litrev}

Bio-inspired optimization algorithms have been studied since the 50's, but they did not gain enough importance till the 90's due to the lack of computers with enough processing capability \cite{Moscato03}. As soon as they were efficiently implemented, results were promising due to the fact that they handle a higher number of variables and achieve better results than traditional methods \cite{Krasnogor06}.

Memetic algorithms are the descendants of genetic algorithms \cite{Moscato03} as they introduce several concepts and modifications in order to obtain nearly optimal results. Standard GAs suffered from several limitations, such as the tendency of getting stuck into local minima and slow convergence speed. In order to tackle these challenges, MAs exploit the concept of {\em meme}, where the best individuals of a certain population share beneficial information about the fittest solutions, allowing even the worst individuals to follow good paths in the search space \cite{Dorigo99}.

However, the tendency of all individuals to reach the same solution lead to local minima and premature convergence problems. In order to avoid premature convergence and enhance global search, memetic algorithms like Particle Swarm Optimization (PSO) \cite{Aggarwal04,Ipadhyaya14} introduced the concept of {\em chaotic movement}. These MAs tackle the premature convergence problem by adding a chaotic component to individual evolution, so the way in which the worst individuals behave is determined by both the best ones and a certain random variable.

Finally, additional information about the search space is used by MAs in order to achieve a faster convergence and greater quality of solutions \cite{Sidhu15}. This is achieved by imposing some restrictions or modifications to the initial solutions, consequently they are not generated in a complete random process but by a pseudorandom procedure, which allows to achieve better solutions.

\section{Theoretical foundation}
\label{sec_theoretical}

\subsection{Digital FIR filter design}

FIR digital filters are defined by their transfer function $H(z)$, which only features the coefficients of the numerator $b_n$ due to the lack of poles, consequently all denominator coefficients are set to one:
\begin{eqnarray}
 H(z) = \sum_{n=0}^{N-1} b_n z^{-n}
\label{ec_1}
\end{eqnarray}
Coefficients $b_n$ in Eq. (\ref{ec_1}) are the coefficients of the numerator and determine the module and phase of each zero in the $Z$-plane and $N$ is the number of coefficients, representing the {\em order} of the digital filter.

The zeros of the transfer function determine the position and magnitude of the minima observed in the module of the frequency response function. The phase of the zeros determines the normalized frequency in which minima are located and the module determines the magnitude.

\subsection{Digital IIR filter design}

Contrary to FIR filters, IIR filters are defined by a transfer function $H(z)$ given by Eq. (\ref{ec_2}), which includes both numerator and denominator coefficients. The first ones determine the zeros while the second ones specify the location of poles.
\begin{eqnarray}
 H(z) = \frac{\sum_{n=0}^{N-1} b_n z^{-n}}{1+\sum_{n=0}^{N-1} a_n z^{-n}}
\label{ec_2}
\end{eqnarray}
In this transfer function, the coefficients of the denominator, $a_n$, determine the phase and module of the poles in the $Z$-plane which represent the maxima in the module of the transfer function.

When designing IIR digital filters, stability must be ensured by limiting the module of the poles by one. As seen in Eq. (\ref{ec_2}), for the same filter order, IIR  digital filters need twice as much coefficients as FIR digital filter, which implies an increase in the computational load of the algorithm (notice that, for simplicity, we have chosen the same order in numerator and denominator; in the general case, orders could be different). However, for the same filter order, IIR filters obtain better frequency response than FIR filters.

\section{Design methodology}
\label{method}

\subsection{Fitness function}
\label{fla}

Shuffled Frog-Leaping Algorithms are based on the behaviour of a pond full of frogs looking for food, each one positioned over a rock whose position determines a solution to the problem. In order to determine the amount of food that a frog finds in a position, the fitness function given by
\begin{eqnarray}
    f(\omega) = \frac{1}{1+J}
\label{ec_3}
\end{eqnarray}
where $J$ is the cost function, determined by the mean square error (MSE) between the desired filter and the current solution. The MSE is calculated using
\begin{eqnarray}
    MSE = \frac{1}{n} \sum_{\omega=0}^{n-1}[h_i(\omega)-h(\omega)]^2
\label{ec_4}
\end{eqnarray}
where $n$ is the number of samples of the obtained filter, $h_i(\omega)$ represents the amplitude of the desired filter and $h(\omega)$ is the amplitude of the obtained filter.

Higher fitness values in Eq. (\ref{ec_3}) correspond to lower MSE values between the desired filter and the obtained one, so the quality of a solution is proportional to its fitness value. Frogs  finding more food are the ones who represent better solutions to the problem, in this case, filters whose response is closer to the ideal one. The best frogs act as leaders for the rest of the population, conditioning the direction in which the other frogs leap.

\subsection{Modeling the solutions}

Each solution from a particular iteration is made up by an array which contains all the filter coefficients, i.e.
\begin{eqnarray}
 w = [a_0, a_1, \ldots , a_n, b_0, b_1, \ldots, b_m]
\label{ec_4b}
\end{eqnarray}
This equation particularly determines the position of a frog when designing IIR filters as two types of coefficients are used. When FIR filters are being designed, only the $b_m$ coefficients are used as the $a_n$ coefficients have zero value, with the exception of $a_0$, whose value is fixed and set to one:
\begin{eqnarray}
 w = [b_0, b_1, \ldots, b_m]
\label{ec_5}
\end{eqnarray}

\subsection{Memeplexe generation}

In order to enhance communication and information interchange between frogs, groups of memetic individuals --or {\em memeplexes}-- are created.

\begin{figure}
  \centering
   \includegraphics[width=0.45\textwidth]{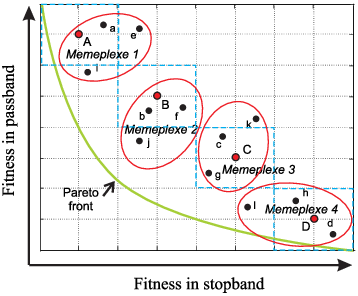}
  \caption{Memeplexe generation and location of each frog in the fitness plane.}
\label{figure_1}
\end{figure}

Each memeplexe should contain the same amount of frogs, so the total population $N$ must be a multiple of the number of memeplexes $M$. These memeplexes are built regarding the characteristics of each frog, so frogs with similar characteristics are grouped together \cite{Luo18}. All frogs are represented in the plane, as shown in figure \ref{figure_1}, where each axis represents the cost function of the resulting filter in both passband and stopband.
The centroid of each cluster is positioned as shown in figure \ref{figure_1}. This way, solutions with similar fitness values are grouped together in the same memplexe. This  technique allows a better evolution of each memeplexe as mentioned in \cite{Luo18}.

\subsection{Adaptation to multiobjective problems}

In this paper a multiobjective problem is tackled, as the intention is to minimize the MSE value of the obtained filter with respect to the ideal one, in both passband and stopband, so two objectives arise.

As two functions are used to evaluate the fitness for each frog, determination of the best and worst frogs is not an obvious task. In this algorithm, the module of the positioning vector of each frog in the fitness plane determines the best and the worst one within each memeplexe. The following equation is used to calculate the module, so the one with the highest module value would be considered as the best frog (the analogous criteria is applied to the worst frog):
{\small
\begin{eqnarray}
 module = \sqrt{(origin_x-fitness_x)^2+(origin_y-fitness_y)^2} \nonumber
    \label{ec_6}
\end{eqnarray}
}%
where subindex $x$ refers to the stopband fitness axis, an the $y$ subindex refers to the passband fitness axis.

\subsection{Leaping process}

Each iteration, $N$ of the worst frogs would leap towards a new position searching for a better solution to the problem \cite{Eusuff04}. First, each frog leaps according to Eq. (\ref{ec_7}), then, the new position is evaluated against the previous position. If an improvement is achieved, the new position is saved:
\begin{eqnarray}
x'_w = x_w +C \cdot rand(0,1)[x_b-x_w]
\label{ec_7}
\end{eqnarray}
where $x_b$ is the position of the best frog of the leaping frog memeplexe, and $x_w$ is the position of the leaping frog. If the new position does not match requirements, a new leap is performed based on the global best frog, according to
\begin{eqnarray}
x'_w = x_w + C \cdot rand(0,1)[x_g-x_w]
\label{ec_8}
\end{eqnarray}
where $x_g$ is the position of the global best frog. In both Eqs. (\ref{ec_7}) and (\ref{ec_8}), a uniform random variable is introduced, denoted as $C \cdot rand(0,1)$, so each leap is scaled by a random number from a uniform distribution in the interval $[0,1]$ multiplied by the factor $C$. This factor decreases linearly from its maximum in the first iteration towards its minimum value in last iteration. This technique allows horizontal (explorative) search at the beginning of the process and a more exploitative search during last iterations \cite{Schonlau98}.
	
The above allows the algorithm to look for solutions in a wider area of the search space at the beginning, and to fine-tune solutions in the last iterations enhancing  local search, so small changes are applied to each frog during the latest iterations \cite{Schonlau98}. Thirdly, if any improvement is made after applying both leaps, a new random frog is generated, which substitutes the previous worst frog as mentioned in \cite{Luo18}.

\subsection{Diversity control}

The lack of diversity of  solutions found during the optimization process can lead to local minima due to premature convergence. To ensure that diversity is maintained along iterations a new entropy-based method is here proposed.

By calculating the Shannon entropy of the vector containing the distances from one frog to the others following Eq. (\ref{ec_9}), diversity can be quantified, so that solutions that get a lower entropy value are usually better distributed (with higher diversity) in the solutions space. In contrast, those sets of potential solutions with a higher entropy value have a more evenly distributed vector of distances, thus lowering the diversity of the population \cite{Laumanns02}. Entropy is calculated as
\begin{eqnarray}
H(x) = -\sum_{i=0}^{n-1}p(x_i)\log _2{p(x_i)}
\label{ec_9}
\end{eqnarray}
where $n$ is the length of the distances vector and $x_i$ are the values of the calculated distances.

To maintain a higher diversity among the first and the second best frogs, the one with a lower entropy value is chosen as the fittest.

\section{Results}
\label{sec_results}

This section shows the results of the numerical simulations carried out with {Matlab\textregistered} and the performance of the proposed filter design method. First, the obtained IIR filter is compared with the one obtained using the BZT  method with elliptic approximation. Then, properties of our designed FIR filters are compared to filters obtained using the windowing method.  Finally, performance of the proposed SFLA is compared to the results reported in \cite{Ipadhyaya16}.

Regarding FIR and IIR design, different filter orders have being used in order to test the evolution of the customized SFLA when parameters are modified. In addition, before any comparison is made, parameter tuning has been carried out in order to obtain the highest fitness value. Optimal parameter values are next shown in Table \ref{tabla_param}.

\begin{table}[h]
\begin{center}
\caption{Best empirical parameter values.}
\begin{tabular}{|c|c|}
\hline
Parameter       &  Value \\
\hline
\hline
    Max. iterations & 500 \\
\hline
    Number of leaps & 8 \\
\hline
    Number of memeplexes & 5 \\
\hline
    Total population & 40 \\
\hline
\end{tabular}
\label{tabla_param}
\end{center}
\end{table}

\vspace{5mm}

\subsection{IIR filter design}

When testing the proposed algorithm, passband and stopband ripples are measured, then an elliptic filter is generated imposing the same values of the mentioned metrics. Finally, a fitness value is calculated for both filters, comparing them to an ideal low-pass filter (LPF), setting $\omega=0.25$ as normalized cut-off frequency. Numerical results are shown in Table \ref{tabla_1}, where all ripple values are given in dB.

\begin{table}[h]
\begin{center}
\caption{IIR design results for proposed SFLA memetic algorithm and BZT with elliptic approximation.}
\begin{tabular}{|c|c|c|c|c|c|}
\hline
SFLA       &  Elliptic  & Passband   &  Stopband    & SFLA       &  Elliptic \\
order    &  order     & ripple     &  ripple      & fitness  &  fitness  \\
\hline
\hline
    5 & 2 & 0.708 & 14.095 & 0.9783 & 0.9750 \\
\hline
    10 & 3 & 0.536 & 21.072 & 0.9918 & 0.9904 \\
\hline
    15 & 4 & 0.73  & 22.2 & 0.9958 & 0.9931 \\
\hline
    20 & 5 & 0.872  & 29,02 & 0.9971 & 0.9947 \\
\hline
    25 & 4 & 0.805 & 23.267 & 0.9979 & 0.9929 \\
\hline
\end{tabular}
\label{tabla_1}
\end{center}
\end{table}

\begin{figure}
  \centering
   \includegraphics[width=0.5\textwidth]{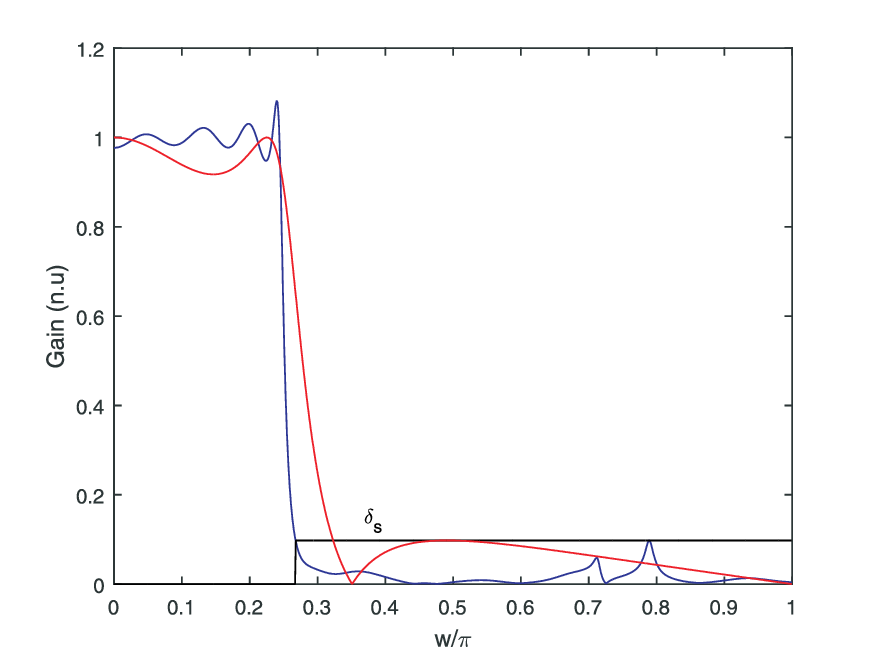}
  \caption{LPF obtained by the proposed SFLA memetic algorithm (blue) and BZT with elliptic approximation (red), for order 25.}
\label{figure_2}
\end{figure}

As shown in Table \ref{tabla_1}, fitness values corresponding to the filters generated by the proposed MA are greater than those using the BZT with the elliptic approximation. However, when using the BZT method, a lower filter order is required and a narrower transition band is obtained.
The reason for the elliptic method to obtain a lower fitness value is a constant stopband and passband ripple, as the filters obtained by the proposed SFLA present a decreasing ripple in both bands as shown in Figure \ref{figure_2}, which is a desired feature in certain applications.

Figure \ref{figure_2} shows the frequency response of the proposed SFLA filter in blue and the elliptic filter in red, proving that the mean ripple in both bands obtained with the proposed method is better. Further testing was carried out using high-pass, band-pass and stop-band filters, obtaining similar results as the ones achieved in the low-pass case.

\subsection{FIR filter design}

For more detailed results regarding performance, FIR LPFs were also generated. To calculate the cut-off frequency $\omega_p$, the passband ripple was established at 1 dB, while ripple in the stopband is set by the filter with the highest value. Table \ref{tabla_fir} shows the fitness values obtained for different filter orders with both the proposed SFLA and the windowing method (with a Blackman-Tuckey window).

\begin{table}[h]
\begin{center}
\caption{FIR design results for SFLA MA and windowing method (Blackman-Tuckey).}
\begin{tabular}{|c|c|c|c|c|c|}
\hline
Filter   &  Stopband     & SFLA        &  Windowed  & SFLA       &  Windowed \\
order    &  attenuation  & cut-off   &  cut-off   & fitness  &  fitness  \\
         &  (dB)         & frequency &  frequency &          &           \\
\hline
\hline
    5 & 15.488 & 0.157 & 0.181 & 0.9702 & 0.9568 \\
\hline
    10 & 20.258 & 0.202 & 0.109 & 0.9829 & 0.9568 \\
\hline
    15 & 17.278 & 0.22 & 0.152 & 0.9876 & 0.9754 \\
\hline
    20 & 16 & 0.232 & 0.203 & 0.9903 & 0.9834 \\
\hline
\end{tabular}
\label{tabla_fir}
\end{center}
\end{table}

Results in Table \ref{tabla_fir} show that fitness values obtained with SFLA are far superior than the ones for the windowing method. However, as the filter order increases, fitness values for both methods tend to match. For further testing, the transition band width was also measured for both methods, showing up the superiority of the proposed method as much lower values were achieved.

\begin{table}[h]
\begin{center}
\caption{Transition bandwidth for proposed SFLA and windowing methods.}
\begin{tabular}{|c|c|c|}
\hline
Order     &  Proposed SFLA     &  Windowing   \\
\hline
\hline
    5 & 0.224 & 0.329 \\
\hline
    10 & 0.143 & 0.263 \\
\hline
    15 & 0.09 & 0.17 \\
\hline
    20 & 0.062 & 0.099 \\
\hline
\end{tabular}
\label{tabla_fir_anchos}
\end{center}
\end{table}

\begin{figure}
  \centering
   \includegraphics[width=0.5\textwidth]{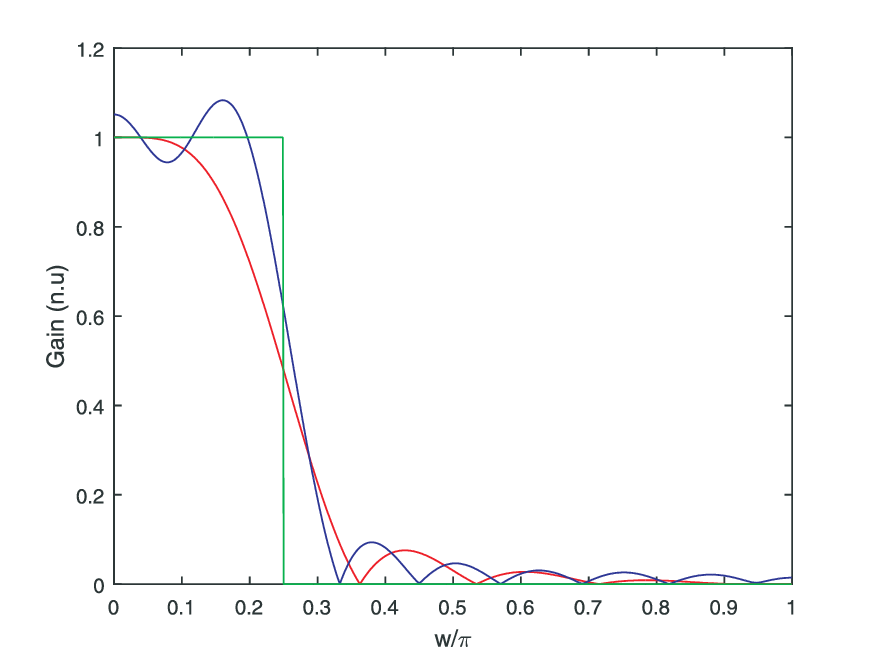}
  \caption{FIR filter comparison between: ideal one (green), proposed SFLA (blue) and windowing (red).}
\label{fir}
\end{figure}

\subsection{Comparison with other optimization algorithms}

In order to compare the obtained results with the ones presented by Upadhyaya et al. in \cite{Ipadhyaya16}, a 5th-order LPF filter was approximated by a 4th-order filter using the proposed memetic algorithm. Eq. (\ref{ec_10}) represents the frequency response of the objective filter.
\begin{equation}
\resizebox{1\hsize}{!}{$
H(z) =\frac{0.1084+0.5419z^{-1}+1.0837z^{-2}+1.0837z^{-3}+0.5419z^{-4}+0.1084z^{-5}}{1+0.9853z^{-1}+0.9738z^{-2}+0.3864z{-3}+0.1112z^{-4}+0.0113z^{-5}}
$}
\label{ec_10}
\end{equation}

For both algorithms, five independent runs were carried out, using the MSE value as indicator. The MSE values for the proposed SFLA and the three bio-inspired methods compared --differential evolution (DE), particle swarm optimization (PSO), and regularized global optimization (RGA)-- are shown in Table \ref{tabla_ffa}.

\begin{table}[h]
\begin{center}
\caption{Minimum Square Error (MSE) values for several compared algorithms.}
\begin{tabular}{|c|c|c|c|c|c|}
\hline
Execution     &    DE        &  PSO      &   RGA       &  Proposed SFLA \\
\hline
\hline
    1  & 6.88E-04 & 0.0277 & 0.0356 & 7.249E-4 \\
    \hline
    2  & 14.0E-04 & 0.0103 & 0.0507 & 5.807E-4 \\
    \hline
    3  & 22.0E-04 & 0.0068 & 0.0991 & 7.4292E-4 \\
    \hline
    4  & 9.72E-04 & 0.0177 & 0.0307 &  10.001E-4 \\
    \hline
    5  & 16.0E-04 & 0.0035 & 0.0556 & 6.3074E-4 \\
\hline
\end{tabular}
\label{tabla_ffa}
\end{center}
\end{table}

As shown in Table \ref{tabla_ffa}, the proposed SFLA MA outperforms all the three algorithms compared, in terms of mean squared error (MSE).

\section{Conclusions}
\label{conclusion}

In this paper we have presented a metaheuristic optimization method based on the standard SFLA with several modifications in order to improve filter design performance of the resulting memetic algorithm. The main aim of the proposed method is the efficient design of digital filters (both FIR and IIR) using two objective functions.

First, a memeplexe generation method, based on the objective function plane, establishes the center of each memetic subgroup in order to improve the convergence speed. Additionally, a mutation mechanism is introduced on the best solutions that come into play when stagnation into  local optima is detected. This way, convergence is maintained and quality of final solution improves. The proposed algorithm has been adapted to multiobjective problems in order to control which characteristics of the solutions are important in the optimization process.

Experimental results are promising as an improvement has been made with regards to solution quality and adaptability to different problems. However, some drawbacks raised in relation to computational load as the proposed SFLA takes more time than traditional methods in order to reach the final solution. Consequently, a trade-off between solution quality and running time must be established by properly adjusting parameters. This situation was observed for IIR design when BZT was used for comparisons. On the other hand, results with our SFLA were better in FIR design whichever method was used for comparison (traditional or bio-inspired).

\bibliographystyle{IEEEtran}
\bibliography{refs}

\end{document}